\documentclass[referee]{aa}   
\begin{document}

   \thesaurus{06     % A&A Section 6: Form. struct. and evolut. of stars
              (03.11.1;  % Cosmogony,
               16.06.1;  % Planets and satellites: general,
               19.06.1;  % Solar system: general,
               19.37.1;  % Stars: formation of,
               19.53.1;  % Stars: oscillations of,
               19.63.1)} % Stars: structure of.
   \title{The effect of dust on photometric redshift measurement;
A self-consistent technique}

   \author{Bahram Mobasher\inst{1}\thanks{Affiliated to the Astrophysics Division of
the European Space Agency}  
\and  Paola Mazzei\inst{2} }  
 \institute  
{Space Telescope Science Institute,
    3700 San Martin Drive,
    Baltimore, MD 21218, USA\\ email: mobasher@stsci.edu \and    
Osservatorio Astronomico,
       Vicolo dell'Osservatorio, 5, 
       35122 Padova, 
       Italy\\ email: mazzei@pdmida.pd.astro.it }
\authorrunning{B. Mobasher \& P. Mazzei}
\titlerunning{A self-consistent photometric redshift technique}  

   \date{ }
\maketitle{}
 
\begin{abstract}
 
A new method is developed for estimating photometric redshifts, 
using realistic template SEDs, extending over four decades in
wavelength (i.e. from 0.05 $\mu m$ to 1 mm). The template SEDs 
are constructed for four different spectral types of galaxies
(elliptical, spiral, irregular and starburst), satisfying the following
characteristics: a). they are normalised to produce the observed colours
of galaxies at $z\sim 0$; b). incorporate the chemo-photometric 
spectral evolution of galaxies of different types, in agreement with
the observations; c). allow treatment of dust contribution and its
evolution with redshift, consistent with the spectral evolution
model; d). include absorption and re-emission of radiation by dust
and hence, realistic estimates of the far-infrared radiation; e). include
correction for inter-galactic absorption by Lyman continuum and Lyman
forest. 
Using these template SEDs, the photometric redshifts are estimated
to an accuracy of $\Delta z = 0.11$. 

The simultaneous and self-consistent modelling of both the photometric
and chemical evolution of galaxies and the effect of dust, makes this
technique particularly useful for high redshift galaxies. 
The effect on the estimated photometric redshifts, due to   
assumptions in the evolutionary population synthesis models, 
are investigated and discussed. Also, the degeneracy in the predicted
photometric redshifts and spectral types are examined, using
a simulated galaxy catalogue.

\end{abstract} 

 
\section{Introduction}

Recent multi-waveband galaxy surveys, carried out from UV to radio 
wavelengths, have identified a population of rapidly star forming galaxies
in the range $0 < z < 3$ 
(Sullivan et al 1999, Lilly et al 1996, Cowie et al 1997, 
Rowan-Robinson et al 1997, Barger et al. 1998; Hughes et al 1998, Mobasher et al 1999). 
In particular, these studies confirm a relatively 
higher rate of star formation in the past, as supported 
by the discovery of a
population of massive, starforming galaxies at $z \sim 3$ 
(Madau et al 1996; Steidel et al 1996).
These objects are likely to be progenitors of the present day galaxies
(Giavalisco et al 1996; Lowenthal et al 1997) and
hence, a statistical study of this population from the epoch of galaxy
formation to the present, gives  clues towards scenarios of the formation and
evolution of galaxies (Fukugita et al 1996). This can also constrain the star
formation history of galaxies out to $z\sim 3$. Such studies require redshift
information for a population of starforming galaxies at faint levels.
Recently, the depth of the available surveys is greatly extended by the 
Hubble Space Telescope (HST)  
observations of the Hubble Deep Fields (HDF). 
The large spectral coverage of the HDFs provide a unique
opportunity to study evolutionary properties of faint galaxies.

Ground-based spectroscopic measurements of the brighter ($I < 25$ mag.)
sub-sample of the HDF have been performed
(Cohen et al 1996; Steidel et al 1996; Lowenthal et al 1997;
Zepf et al 1996). However, for the fainter galaxies
in the HDF, redshift measurements are more difficult with spectroscopic
features almost impossible to identify.
For these objects, the
photometric redshift technique (Loh and Spillar 1986; Connolly et
al 1995) is faster than its spectroscopic counterpart and applicable to much
fainter magnitudes. This is due to a larger bin size in photometry compared to
spectroscopy ($\sim 1000$ \AA\ $vs.\ 1-2 $\AA), leading to a
shorter exposure time
with a trade-off in accuracy of the measured redshifts. 

Considering the new generation of 8m class telescopes, the planned 
instrumentation on the HST and future, high sensitive radio
telescopes, a substantial number of deep surveys at different wavelengths
will soon become available. Most of these galaxies will be too faint for
spectroscopic study and hence, the photometric redshift technique is the
only practical way for estimating their redshifts. 
In a recent assessment of different photometric redshift techniques, using
a redshift-limited spectroscopic survey, it was shown that photometric
redshifts can be predicted with an accuracy of 0.1 (0.3) for $68\%$ ($99\%$)
of the sources examined (Hogg et al 1998). Therefore, photometric redshifts
could provide a powerful tool for {\it statistical} studies of
evolutionary properties of galaxies and in particular of
faint galaxies for which spectroscopic data are difficult to obtain.

The most important step in any study concerning photometric redshift 
measurement, is the choice 
of the template Spectral Energy Distributions (SEDs) for different
populations of galaxies, with which the observed SEDs should be compared. 
There are two general ways for adopting the template SEDs:

a). Empirical templates: in this case one uses the mean {\it observed} SEDs 
corresponding to 
different types of galaxies. The problem here is that there are not
enough information about the observed SEDs for different classes of objects at
different redshifts (particularly at high redshifts). Therefore, 
incorporating the spectral evolution of galaxies of different types on their 
template SEDs is difficult and uncertain. 

b). Synthetic templates: uses model SEDs for different spectral types
of galaxies, shifted in redshift space, assuming evolutionary population
synthesis models. The main problem here is to constrain the evolutionary model
parameters to produce realistic model SEDs (for different types) 
as a function of redshift. In particular, the effect of dust at high redshifts
(specially in star forming galaxies) is not known.

To avoid these problems, we introduce a combined approach, 
producing realistic model SEDs based on
chemo-photometric Evolutionary Population Synthesis (EPS) models, extending
from UV to 1 mm in wavelength. The template SEDs here, 
are consistently and simultaneously optimised to;  
a). produce the observed colours of galaxies at $z\sim 0$; 
b). incorporate chemo-photometric evolution for 
galaxies of different types, in agreement with observations;
c). allow treatment of dust contribution 
and its evolution with redshift, consistent with the EPS models; 
d). include absorption and re-emission of radiation by dust
and hence, realistic estimates of the far-infrared radiation;
e). include correction for inter-galactic absorption by Lyman continuum and
Lyman forest. {\it The evolutionary models and hence, the template SEDs, 
are constrained by minimising the scatter
between the photometric and spectroscopic redshifts for a calibrating
sample of galaxies with known spectroscopic data}.

The main advantage of this technique over the previous works
is that it simultaneously and
self-consistently allows for the treatment of both the photometric and
chemical evolution of individual galaxies with time.  
Moreover, since the synthetic template SEDs cover the range
from UV to sub-mm wavelengths, one could consistently use the 
optimised SEDs to estimate contributions from individual galaxies
to the far-infrared and sub-mm wavelengths. Also, the effect of dust and its
evolution with redshift is self-consistently accounted for in the template
SEDs and optimised to produce the observations. This is crucial in any
photometric redshift technique if it is to be applied on high redshift, 
star-forming galaxies (Meuer et al 1997; Cimatti et al 1998). 

The new photometric redshift technique is outlined in the next section. 
In section 3, the evolutionary population synthesis models are briefly
discussed. Section 4 presents 
the calibration sample. This is followed by the optimised template SEDs in
section 5. The uncertainties in the photometric redshifts and spectral 
types are explored in section 6. The conclusions are presented in section 7. 

\section{Photometric Redshift Technique}

For a given galaxy, the photometric redshift can be estimated by comparing its
observed SED with a set of template SEDs, corresponding to 
different morphological types and shifted to different redshifts, accounting
for galaxy evolution with look-back time. 
The redshift and spectral type associated with the template SED closest to the
observed SED will then be assigned to that galaxy. The comparison is 
carried out by minimising the $\chi^2$ function 
 
$$\chi^2 = \sum_{i=1}^n ((F_{obs}^i - \alpha F_{template}^i)/\sigma^i)^2$$
where the summation, $i$, is over the passbands with
$F_{obs}^i$ and $F_{template}^i$ being, respectively, the observed and
template fluxes at a given passband, $n$, $\sigma^i$ the
uncertainty in the observed flux and $\alpha$ the normalisation, 
estimated in two different ways.  
First, it is estimated by forcing the template SED to have the same total
energy (calculated by integrating the
SED over the observed wavelength range) as the observed SED. 
Second, the normalisation is used as a free parameter in the fit in minimising
the $\chi^2$. The two methods give very close results. 
The comparison between the observed and model SEDs,   
over the entire range of observed SEDs, will     
constrain the three free parameters (i.e. redshift, spectral type
and normalisation). However, due to changes with redshift in the
properties of galaxies, we also need to allow for evolutionary effects
on the model SEDs. This introduces more free parameters which will be 
constrained using observed data, as discussed in the following sections. 
  
In order to construct the template SEDs needed for photometric redshift
measurements, we use Evolutionary Population Synthesis (EPS) models, 
required to
estimate changes in  properties of galaxies with redshift. These models 
allow us to develope the template SEDs as a function of redshift for different
types of galaxies. In the next section, the EPS models used to
derive the template SEDs for different types of galaxies, and their underlying
assumptions are discussed. The models are constrained, using observed
{\it local} SEDs for different types of galaxies 
and by optimising the estimated
photometric redshifts to produce their spectroscopic counterparts for a 
calibrating sample
of HDF galaxies with known spectroscopic redshifts. 
Details of the procedure adopted in this study are summarised in the flow chart
in Figure 1.  

\section {The Evolutionary Population Synthesis Models  }

Synthetic SEDs are constructed following a self consistent procedure,
incorporating stellar emission, internal extinction and re-emission by dust.
These chemo--photometric EPS models, providing the SEDs extending
over four decades in wavelength (ie. from 0.05$\mu\,m$ to 1 mm)-
(Mazzei et al 1992, hereafter MXD92), allow us to investigate 
the local properties and the evolution with galactic age of the SEDs of 
different galaxy types. In particular, we use realistic EPS models
for
disc galaxies (MXD92), early--type galaxies (Mazzei, De Zotti and Xu (1994),
hereafter MDX94; 
Mazzei and De Zotti (1996)) and the far-infrared luminous starburst 
population (Mazzei, Curir and Bonoli (1995)).  

In order to match the photometric properties of the galaxy types, 
expected to dominate deep surveys, 
we consider 4 different templates, consisting of; elliptical, spiral,
irregular and starburst. Details of the SED models for these types and their
evolution with cosmic time are given in the above references. In this section, 
an overview of the main parameters (to be optimised using observed SEDs) 
will be presented. 

The ESP models, used to generate synthetic SEDs for different 
types of galaxies, have been linked to their chemical evolution so that
the increased metallicity of successive stellar generations is taken into
account. We adopt a parametrization for the time
evolution of the star-formation rate as 
$$\psi (t) = \psi _0 f_g^n(t)\ M_\odot/yr,
\eqno(1) $$ where
$f_g$ is the fractional mass of gas which takes part in the star formation 
($f_g =m_{gas}/m_{gal}$)
and $\psi_0$ is the initial star-formation rate ( i.e. SFR at $z=z_{form}$; 
where $z_{form}$ is the formation redshift). We assume that initially
$f_g = 1$ with $m_{gal} =10^{11}$ M$_\odot$. 

The synthetic spectra for the stellar generations  with different metallicities
which contribute to the galaxy SED
are estimated at any given time, using the recent set of  isochrones
constructed by Bertelli et al. (1994). These incorporate the results of new
stellar evolutionary calculations, based on six metallicity values:
0.0004,0.001,0.004,0.008,0.02 and 0.05. The isochrones  have a fine coverage
of masses and ages and include almost all the evolutionary phases from the
main sequence to the stage of planetary nebulae ejection or carbon ignition, as
appropriate for a given initial mass. This allows us to link the photometric
and chemical evolution in galaxies.
Further, at any given time and wavelength, 
the models incorporate both the average correction due
to internal extinction by the enriched ISM, assuming  a dust--to--gas ratio
proportional to the gas metallicity and the  dust re-radiation from 3
$\mu$m up to 1 mm.
We assume that gas and stars have the same distribution. In particular,  
for spiral, starburst and irregular galaxies we follow the same approach 
as described in MXD92
(i.e.  an exponential function of the galacto-centric radius- 
see their eq.(11))
whereas for ellipticals we follow MDX94 (i.e a density profile given by 
the King (1966) model). Results are not strongly dependent on the adopted 
geometry. 
The evolution of the optical depth, $\tau$, follows directly from that of the
gas metallicity, $Z_{gas}$ and $f_g$  since $\tau (t)\propto Z_{gas}(t)^s
f_{g}$ ( as a consequence of the assumption of a gas-to-dust ratio being 
proportional to the gas metallicity; MXD92, MDX94).  
This results in very different histories for the effective optical depths
(where $L_{FIR}/L_{bol}=1-exp(-\tau_{eff}$))   
in the early phases of the evolution
of ellipticals in the model, which is based on the adopted IMF parameters. 

The IMF and the SFR are the main parameters, controling the $e$-folding star
formation time scale, the chemical evolution and, as a direct consequence, the
optical depth of the system. Thus, the far-infrared data, where available, are
expected to constrain the galaxy metallicity. Therefore, by comparing the local
observed SEDs of different galaxy types with models, we constrain both
$\psi_0$ and the IMF.
The observed SEDs for nearby
spiral and elliptical galaxies with known metallicities and far-IR (IRAS)
observations are used to constrain the IMF parameters so that 
these models
successfully reproduce the local chemo--photometric properties of both
these types over four decades in wavelength.
However, the parameter which is most sensitive to the local properties
(i.e. the gas content and optical colours) of galaxies along the
Hubble sequence, is the initial star formation rate, $\psi_0$
(Sandage 1986), which will be constrained for different types of
galaxies, using the observed SEDs (section 5.1). 

\section {The Calibration Sample}

To further constrain the evolutionary parameters in these models, 
in particular $n$ and
$z_{form}$, a calibration sample of HDF galaxies with available spectroscopic
redshifts is compiled.
The calibration sample contains galaxies bright enough to allow spectroscopic
redshift measurements, consisting of both objects with UBVI detections and the
UV `drop outs' (i.e. high redshift objects).  
The advantage of using the HDF galaxies as the calibration
sample here is two-fold: (a). HDF galaxies cover the spectral range
required for photometric redshift measurement, containing galaxies
in the range $0 < z < 3.5$; (b). they are selected using a uniform criteria,
for both nearby and distant galaxies, making this an unbiased
calibration sample. 

The calibrating galaxies are
individually inspected and for galaxies with close neighbours, a smaller
aperture is adopted to avoid contamination of their light by nearby objects.
For the objects with no UV detection (i.e. UV `drop outs'), a U-band
magnitude of 28.01 mag. was assumed for the photometric redshift measurement. 
Although this is likely to introduce a bias due to a colour-magnitude 
relation (i.e. fainter galaxies are bluer), we do not expect it to be 
significant. This is examined by exploring the range $28.0 < U < 30.0$
for each of the galaxies in the UV- drop out sample and estimating
their respective photometric redshift. Using models with the same parameters, 
we find, on average, only a small sensitivity  
of the estimated photometric redshifts ($ < 3\sigma$) 
to changes in the UV magnitudes in the above range. 

The calibrating sample, consisting of 73 galaxies, is listed in Table 1
together with their UBVI photometry and spectroscopic redshifts. 
 The reliability of the spectroscopic
redshifts and the photometric accuracy of individual galaxies are discussed in
the footnote to this table. The calibration sample in Table 1 was selected to
be in the HDF area, to allow accurate UBVI photometry and to have unambigious
spectroscopic redshifts. Moreover, objects with non-stellar sources of energy
(i.e. gravitationally lensed candidates; Zepf et al (1996)) are not included.
The magnitudes are in the AB system and are measured over an aperture of 3 
arcsec
diameter (unless stated otherwise in the footnotes to Table 1). 

The EPS models for different types of galaxies are therefore further 
constrained by minimizing the {\it rms} scatter between the photometric
redshifts, predicted by our models, and their spectroscopic counterpart, using
the calibration sample in Table 1. Thus, the final EPS models for different
types of galaxies, which we define as templates (see below), have local
($z=0$) SEDs which match the local, observed, SEDs of the corresponding galaxy
types  and  evolutionary properties constrained, using 
the calibration sample here. 

\section  {The Template SEDs}

\subsection{EPS Model Parameters}

A large number of EPS models with different input parameters (IMF, $\psi_0$,
$n$ and $z_{form}$) are developed, each accounting for both the local
properties and evolutionary behaviour of the 4 types of galaxies considered
here. For a given spectral type, the model parameters are normalised at $z=0$
by fitting them to the local observed SED of their respective type.
The evolutionary behaviour of the EPS models are then constrained by 
estimating the photometric redshifts to our calibrating sample in Table 1, 
considering template SEDs with different parameters (i.e. IMF and $z_{form}$)
and allowing for Lyman continuum and Lyman forest absorption for $z > 2$
galaxies.  
The template SEDs corresponding to the evolutionary parameters which 
give the closest agreement between the photometric and 
spectroscopic redshifts are then adopted (see below).  
The sensitivity of the final results (i.e. photometric redshifts) to the 
input parameters in the EPS models is studied in the next section while,  
details of the final templates for individual types, which best 
satisfy the above requirements, are summarised below:

\noindent {\bf a). Ellipticals.}
The synthetic SEDs, representing elliptical galaxies, are constructed taking
$n=0.5$ and $\psi_0=100\,M_\odot/yr$. This gives an $e$-folding
star formation time scale of 1 Gyr and 
reproduces the observed local SEDs for the ellipticals (Figure 2.1a). 
A formation redshift of 5 is estimated, corresponding to an age of 13 Gyrs. 
These models predict a noticeable extinction by dust (A$_B< 4.6$ mag) in the
first stages of their evolution which are characterized by intense star
formation activity,  making them powerful far-infrared sources (Fig 2.2a)
(see  MDX94 for more details). 

\noindent {\bf b). Spirals.}
The models used to produce synthetic SEDs for spiral galaxies have $n=2$ and
$\psi_0=10\,M_{\odot}$/yr. They are consistent with an e-folding star
formation time scale of 10 Gyrs, corresponding to a formation redshift of 2 
(i.e. an age of 10-11 Gyrs). This gives a local value of $L_{FIR}/L_{BOL} =
0.3$, very similar to that of our own Galaxy. This model re-produces the
local SED of NGC 3627  out to the far-infrared wavelengths (Figure 2.1c). 
Spiral galaxies are slowly evolving with time ($L_{FIR}/L_{BOL} < 0.1$ 
beyond $z = 1$), resulting a smooth 
evolution for their SED (Figure 2.2c)-
(see MXD92 for more details).

\noindent{\bf c). Starbursts.}
The template SEDs for the starburst population are produced taking $n=-1$,
$\psi_0=2\,M_{\odot}/yr$ and a formation redshift of 5. This template does not
produce powerful far-infrared emission at any redshift and hence, its local
SED (Figure 2.1b) is different from that of local luminous far-infrared
starbursts (ie. M82 and Arp 220). The SFR in this model is a gradual process
with a smooth time scale, leading to formation of very blue, metal poor
systems at $z\sim 0$ (Figure 2.1b) and a blue, dust-free 
($L_{FIR}/L_{BOL} < 0.05$) system
at $z > 1$ (Figure 2.2b). This mimics a scenario involving frequent but 
short bursts of star
formation, which use a small fraction of gas in these systems (strong bursts
of star formation rapidly exhaust the gas, leading to ellitpical like systems).
The starburst templates here represent the average evolutionary behaviour 
expected for this population of galaxies. 
 
\noindent{\bf d). Irregulars.}
The Irregular template has been produced with the same recipe
as the starburst but with a formation redshift of $z_{form}=1$ (ie. an age of
0.8-0.9 Gyr).
The local SED of the irregular galaxy NGC 4449 (Kennicutt, 1992) is compared in
Figure 2.1d with our local templates for both irregular and 
starburst galaxies. 
These templates are also compared at $z=0.975$
(Figure 2.2d), showing a significant difference beyond $\lambda\sim 1 \mu m$. 

The model SEDs at $z=0$ for the four galaxy types discussed above,
agree well with the local observed SEDs, as shown in Figures 2.1a-2.1d. 
The effects of Lyman break and Lyman forest opacities are included to the
template SEDs, using the relations $\Delta (M)$ vs. $z$ for
different wavelengths, given in 
Madau et al (1996).
These relations were fitted to parametric forms, 
which were then used to
estimate the respective correction (due to absorption by inter galactic
medium) to the SEDs at any given redshift.  
The correction due to Lyman break and  Lyman forest absorption ranges
from $\Delta (M_{UV}) = 0.25$ mag. at $z=2$ to $\Delta (M_{UV}) \sim 1$ mag.
at $z=2.5$. 

\subsection {Sensitivity of the Template SEDs on the Model Parameters}

In this section we study the dependence of the results
(i.e. the photometric redshifts) to the model parameters which 
most strongly affect the final template SEDs and hence, the
predicted photometric redshifts.
These parameters consist of the shape of the IMF and its lower mass limit, the
total number of templates (i.e. spectral types) and the 
formation redshift for each galaxy type.
New template SEDs are generated corresponding to EPS models for 
elliptical, spiral, starburst and irregular
galaxies, using the parameters listed in Table 2, with the rest of the 
parameters taken to be the same as discussed in section 5.1. 
For each set of the new templates, the photometric redshifts are estimated
for galaxies in the calibrating sample (Table 1) with the {\it rms} scatter
in the quantity $(z_{phot} - z_{spec})/(1+z_{spec})$  
(ie. between the photometric and spectroscopic redshifts) calculated and listed
in Table 2. 
The templates corresponding to the model which gives the smallest {\it rms}
estimate in Table 2 (i.e. model 4) is then adopted. 
The photometric redshifts
estimated for the calibration sample, using the adopted template SEDs
(model 4 in Table 2), are listed in Table 1 and compared
with their spectrosopic counterparts in Figure 3. The 
{\it rms} scatter of 0.11 here is taken as 
the uncertainty in the photometric redshift estimates in the range 
$0 < z < 3.5$. 
The uncertainties in correcting for intergalactic absorption at $z > 2$  
indirectly affect the optimisation of the
EPS model parameters in this section. To explore this, we constrained the
calibration sample {\it only} to galaxies with $ z < 2$ 
(which are much less affected by the
IGM absorption) and estimated the EPS model parameters so that to minimise
the {\it rms} scatter in Figure 3. No change is found in the EPS model 
paremeters. Also, we explored the sensitivity of the {\it rms} scatter
in Figure 3 to the number of template SEDs used (ie. including templates
for different sub-classes of spirals), taking the 
number of SEDs as a free parameter. This did not reduce the
optimum {\it rms} scatter, derived using the four templates.     

Considering the results in Table 2, it seems that one set of models
(Salpeter IMF with $m_l=0.01$), give a considerably better fit
to observations. This was extensively tested by exploring the parameter
space, consisting of the IMF shape and its mass limits, 
total number of templates and formation redshifts. 
It was found that this is not due to sampling a particular region
of the parameter space or the number of templates, and is indeed, 
a real effect. A similar study, using
a different set of EPS models and the optimised parameters in Table 2, would be
extremely valuable. 

The galaxies in the calibration sample in Table 1 
also have near-IR data (Fern\'andez-Soto et al. 1998).
The above procedure was repeated, using the combined
optical and near-IR magnitudes (UBVIJK) for the calibrating sample. 
This did not change the optimized EPS model parameters in section 5.1 and
Table 2. 
As an independent test of the template SEDs here, we include the
near-IR magnitudes to the observed
SEDs of the calibrating
sample in Table 1 and estimate their photometric redshifts, using the template
SEDs predicted in section 5.1 (model 4 in Table 2).   
Compared to their spectroscopic 
counterparts, an {\it rms} scatter of 0.13 is found, in agreement
with 0.11 from Figure 3.
 
The amount of dust and its evolution with redshift is an important
characteristic of the EPS models in this study. This, at any time, is
computed self--consistently, accounting for both the SFR and the IMF 
parameters.  
By extending the IMF to low $m_l$ values ($m_l=0.01$ M$_\odot$), 
the gas depletion rate 
becomes faster, reducing the dust enrichment rate. This 
leads to higher optical depth in the early evolutionary phase of our
elliptical models (templates)-(see also Mazzei \& DeZotti 1996 for
more details). 
The UV extinction ($A_U$) corresponding to different IMF and $m_l$ values is
estimated for both elliptical and spiral templates at different redshifts and
are listed in Table 3. 
By changing the shape of the IMF and its lower mass limit,
the UV extinction for ellipticals at $z\sim 2$ 
changes in the range $\sim 1-4$ mag.
Considering the optimised model in Table 2 (model 4), the extinction in
spirals is significantly larger than in ellipticals at $z\sim 0$ while, 
at higher redshifts ($z\sim 1.5$), ellipticals are obscured more than the
spirals. Using our optimised model in section 5.1, 
the template SEDs at $z\sim 1.5$ are compared in Figure 2.2 
with their counterparts at $z\sim 0$. 
These show a significant dust contribution
to the elliptical SEDs at $z\sim 1.5$, 
as indicated from the peak at the far-IR wavelengths.  
 
\section {Uncertainties in the Photometric Redshifts and Spectral Types}

A potential source of uncertainty in estimating the photometric redshifts and
spectral types of galaxies is the possibility that the models might be
degenerate (i.e. different synthetic SEDs, corresponding to different
redshifts and galaxy types producing the same result).
Furthermore, the photometric errors 
in the observed SEDs are likely to affect the final estimate
of both the photometric redshifts and the spectral types of their respective 
galaxy. Also, due to the relative similarity of the model SEDs for
starburst and irregular galaxies (Figure 2), the accuracy with which the
spectral types for these systems are predicted, needs to be established.  

To investigate the above problems, a Monte Carlo simulation is performed. A
simulated catalogue is generated to resemble the observed 
HDF survey, with
UBVI magnitudes, known redshifts ($z_{input}$) and spectral types.
The galaxies are randomly selected to have SEDs similar to the synthetic SEDs
for the four types of galaxies (elliptical, spiral, starburst and irregular)
considered in section 5.1, shifted in redshift space. Random Gaussian noise,
resembling photometric errors, are then added to the simulated SEDs.
The simulated catalogue  
has a magnitude limit of I=28 mag and  
an apparent magnitude distribution similar to the
observed HDF survey.    

The photometric redshift code is used to predict the redshifts 
($z_{output}$) and spectral
types of individual galaxies in the simulated catalogue and to compare them
with their input values. Two sets of simulations is carried out,
assuming different observational errors (i.e. $\sigma$ in the Gaussian noise
distribution).
The difference between the input (simulated) and the output (predicted) 
redshifts is presented in
the histogram in Figure 4, with the spectral types compared in Table 4.

The $log(z_{output}/z_{input}) $ distribution (Figure 4), 
including {\it all} the four types
of galaxies, shows a distinct peak at zero, indicating that the redshifts for
the simulated galaxies are well re-produced within $\Delta z \sim 0.11$. 
The objects, located at the tails of the distribution in Figure 4, 
are all galaxies classified as spirals, irregulars or starbursts
for which, the degeneracy of their SEDs appears to be more serious 
(see below). When increasing the photometric errors
in the simulated SEDs, the agreement between the simulated (input)
and predicted (output) redshifts decreases but 
the distribution still peaks around zero (dotted line in Figure 4). 

The spectral types of the simulated galaxies are compared with the predicted
types in Table 4. This gives the ratio of galaxies of a given spectral type, 
which are correctly classified in the simulation (i.e. galaxies for which
their spectral type in the input catalogue were successfully re-produced), 
to the total number of galaxies of the same type in the input catalogue. 
In both simulations, we re-produce the spectral types for
{\it all} the ellipticals in the input catalogue with no mis-identifications. 
However, due to the relative similarity of the 
UV-to-optical part of the synthetic SEDs for 
the spirals, irregulars and starburst galaxies (Figure 2), we can recover
respectively $79\%$, $85\%$ and $71\%$ of the spectral types
of galaxies in the input catalogue. 

The conclusion from the simulation here is that the 
photometric redshifts are, on average, well produced (within the expected
accuracy) and are not affected by the
degeneracy in the template SEDs. Furthermore, the spectroscopic
type classification for ellipticals is reliable with 100\% re-produced 
(i.e. no mis-identifications) from the input catalogue while, for the
spirals, irregulars and starbursts, there is a slight degeneracy in
predicting their spectral types. 

\section{Summary and Conclusion}

A new technique is developed for estimating the photometric redshifts to
galaxies. 
The advantage of this over the previous methods is that it allows
a self-cosistent treatment of both photometric 
and chemical evolution of the template SEDs for different types
of galaxies, in agreement with observations. 
For this reason, the present technique is
particularly useful for application to high-redshift galaxies. 
The template SEDs are 
constrained to simultaneously satisfy the following criteria:
a). produce the observed colours of galaxies at $z=0$;
b). incorporate the spectral evolution for galaxies of different types;
c). allow a self-consistent treatment of the dust contribution 
and its evolution with redshift;
d). include correction for inter-galactic absorption by Lyman continuum and
Lyman forest.  
The model parameters are then constrained, using a 
a calibrating sample of HDF galaxies. Using {\it } only four optical passbands
(UVRI), the photometric redshifts to galaxies can be estimated with an 
{\it rms} accuracy of 0.11. Including near-IR data (UVRIJK), gives a similar
{\it rms} scatter in the photometric redshifts from this method. 
The sensitivity of the results to different EPS model parameters are explored
and the degeneracy of the photometric redshifts and spectral types for
galaxies are examined using a simulated catalogue.

\clearpage

\begin{figure} 
\caption[fig1.eps]{
Flow chart summarizing the photometric redshift technique. 
}
\label{fig1}
\end{figure}

\begin{figure} 
\caption[FIG1.eps]{ 
Fig. 2.1). 
The synthetic Spectral Energy Distributions (SEDs) are predicted
and constrained to fit the observed data at $z\sim 0$, 
as explained in the text (left panel). Details of each panel and 
the source of the observational data for each type is given below.

a).
Elliptical galaxies: open squares  (Burstein et al. (1988)); 
filled triangles (Schild and Oke (1971)); filled circles 
(Oke and Sandage (1968)); asterix (Kennicutt (1992));
the filled circles at FIR wavelengths correspond to the average local 
FIR SED for these galaxies (Mazzei and DeZotti (1994b)).
  
b).
Starburst galaxies: this corresponds to the observed SEDs of  
NGC5996
(MK691). The observed data are taken from:  
filled circles (Kennicutt (1992)); asterix in the shorter wavelength
region (IUE data from Kinney et al. (1993)); asterix in
the near--IR region (Balzano (1983)); FIR data 
(IRAS Catalugue Version 2 (Fullmer, L. and Lonsdale, C. 1989));
the IUE data are measured over $20''\times 10''$ aperture 
and are shifted vertically by a
factor of 1.6 to normalise to the optical SED; near-IR data, 
measured over $10.3''$ aperture has been shifted by a factor of 4.5.

c).Spiral  galaxies:  this
corresponds to the NGC3627 galaxy. The observed data are from 
Kennicutt (1992) and Rice et al. (1988). 

d). The local observed SED of IRR galaxy NGC 4449 (Kennicutt, 1992), filled
circles, is compared with the synthethic SED of our Irr template at $z=0$
(continuous line) and that for the starburst galaxies (dashed line). 

Fig. 2.2a-c. 
The local synthetic SEDs for different types of galaxies 
from Figures 2.1a-c (solid lines) are compared with their
counterparts at $z\sim 1.75$ (dashed lines). 

The SEDs for
irregulars (solid line) and starbursts (dashed line), both at $z=0.975$, are
compared in Figure 2.2d.

\label{fig1}}
\end{figure} 

\begin{figure} 
\caption[FIG2.eps]{ 
Comparison between the spectroscopic redshifts
with the photometric values estimated here for 73
HDF galaxies from Table 1. The spectroscopic redshifts are 
taken from Cohen et al (1996); Steidel et al (1996); Lowenthal et al (1997);
Dickinson (1998). 
The UV drop-out objects for which a U-band limiting magnitude of 28.01
is used to estimate the photometric redshifts, are also included. 
The crosses are HDF3646+1408
and HDF3659+1222 galaxies which have uncertain published spectroscopic redshifts 
of $z=0.13$ and 0.47 respectively. 
The lines have a slope of unity. 
The photometric redshifts are based on model 4 in Table 2. 
The rms scatter is estimated in $(z_{phot} - z_{spec})/(1+z_{spec})$
and corresponds to 0.11 with
the dashed lines corresponding to 
$\pm 3\sigma$ error. 
\label{fig2}}
\end{figure} 

\begin{figure} 
\caption[FIG5.eps]{ 
Histogram of $log(z_{output}/z_{input})$ for the simulated UV selected HDF 
catalogue. The continous line is the result for the photometric errors
($\Delta U$, $\Delta B$, $\Delta V$, $\Delta I$) corresponding to
(0.15,0.10,0.05,0.05) and the dotted line corresponds to
(0.25,0.15,0.10,0.10). Both the histograms peak around zero, implying that
the redshifts in the simulated catalogue can be produced to within
$\Delta z = 0.11$.
\label{fig3}}
\end{figure}

\begin{table}
\caption{The Calibrating Sample}  
\begin{tabular}{lllllllll}   

    &RA(J2000) & 
Dec(J2000) &
I & R &B & U & 
$z_{spec}$ & $z_{phot}$ \\  
               &               &                &       &       &       &  & &\\  
  HDF3646+1143 & 12\ 36\ 46.12 & 62\ 11\  43.38 & 21.47 & 22.44 & 22.92 & 23.25 
& 1.012 & 0.775 \\ 
  HDF3643+1143 & 12\ 36\ 43.72 & 62\ 11\  43.95 & 20.88 & 22.49 & 24.62 & 24.62 
& 0.764 & 0.700 \\ 
  HDF3646+1146 & 12\ 36\ 46.84 & 62\ 11\  46.09 & 23.50 & 24.17 & 24.24 & 23.89 
& 1.059 & 0.950 \\ 
  HDF3648+1147 & 12\ 36\ 48.31 & 62\ 11\  47.47 & 25.12 & 25.28 & 26.04 & 28.01 
& 2.980 & 3.000 \\ 
  HDF3646+1152 & 12\ 36\ 46.44 & 62\ 11\  52.31 & 21.91 & 22.99 & 24.70 & 24.58 
& 0.504 & 0.425 \\ 
  HDF3643+1152 & 12\ 36\ 43.32 & 62\ 11\  52.63 & 22.91 & 23.33 & 23.42 & 23.61 
& 1.242 & 1.600 \\ 
  HDF3649+1156 & 12\ 36\ 49.26 & 62\ 11\  56.08 & 23.48 & 23.92 & 23.80 & 23.48 
& 0.961 & 0.975 \\ 
  HDF3641+1201 & 12\ 36\ 41.51 & 62\ 12\   1.41 & 24.64 & 25.03 & 25.83 & 25.47 
& 0.483 & 0.650 \\ 
  HDF3645+1202 & 12\ 36\ 45.86 & 62\ 12\   2.63 & 23.66 & 24.44 & 24.83 & 24.26 
& 0.679 & 0.900 \\ 
  HDF3640+1203 & 12\ 36\ 40.70 & 62\ 12\   3.84 & 23.10 & 23.73 & 23.97 & 24.44 
& 1.010 & 1.350 \\ 
  HDF3641+1206 & 12\ 36\ 41.85 & 62\ 12\   6.59 & 21.01 & 21.64 & 22.57 & 23.19 
& 0.432 & 0.600 \\ 
  HDF3651+1210 & 12\ 36\ 51.94 & 62\ 12\  10.69 & 22.92 & 23.44 & 24.17 & 24.81 
& 0.456 & 0.525 \\ 
  HDF3648+1215 & 12\ 36\ 48.17 & 62\ 12\  15.04 & 22.56 & 23.40 & 23.75 & 23.58 
& 0.960 & 0.875 \\ 
  HDF3642+1217 & 12\ 36\ 42.78 & 62\ 12\  17.48 & 20.73 & 21.30 & 22.20 & 22.88 
& 0.454 & 0.550 \\ 
  HDF3643+1219 & 12\ 36\ 43.53 & 62\ 12\  19.33 & 22.59 & 23.42 & 23.92 & 24.71 
& 0.750 & 0.725 \\ 
  HDF3652+1220 & 12\ 36\ 52.60 & 62\ 12\  20.77 & 23.15 & 23.63 & 24.37 & 25.12 
& 0.401 & 0.480 \\ 
  HDF3649+1221 & 12\ 36\ 49.44 & 62\ 12\  21.17 & 24.04 & 24.50 & 24.65 & 24.45 
& 0.961 & 0.950 \\ 
  HDF3651+1221 & 12\ 36\ 51.62 & 62\ 12\  21.30 & 21.54 & 22.23 & 23.32 & 24.61 
& 0.299 & 0.460 \\ 
  HDF3659+1222$^4$ & 12\ 36\ 59.45 & 62\ 12\  22.20 & 23.67 & 24.30 & 24.68 & 24.18 
& 0.470 & 0.925 \\ 
  HDF3648+1222 & 12\ 36\ 48.95 & 62\ 12\  22.38 & 22.43 & 23.21 & 23.53 & 23.50 
& 0.953 & 0.875 \\ 
  HDF3658+1222 & 12\ 36\ 58.56 & 62\ 12\  22.76 & 23.34 & 24.03 & 24.40 & 23.90 
& 0.681 & 0.925 \\ 
  HDF3657+1226 & 12\ 36\ 57.11 & 62\ 12\  26.68 & 22.45 & 23.03 & 23.64 & 23.42 
& 0.561 & 0.700 \\ 
  HDF3651+1228 & 12\ 36\ 51.26 & 62\ 12\  28.47 & 26.33 & 25.94 & 26.41 & 27.30 
& 2.775 & 2.800 \\ 
  HDF3644+1228 & 12\ 36\ 44.55 & 62\ 12\  28.51 & 23.59 & 23.96 & 24.10 & 25.37 
& 2.268 & 2.050 \\ 
  HDF3647+1231 & 12\ 36\ 47.30 & 62\ 12\  31.76 & 22.82 & 23.23 & 23.99 & 23.95 
& 0.421 & 0.550 \\ 
  HDF3653+1235 & 12\ 36\ 53.37 & 62\ 12\  35.41 & 22.81 & 23.48 & 24.21 & 26.13 
& 0.559 & 0.430 \\ 
  HDF3700+1235 & 12  37\  0.48 & 62\ 12\  35.77 & 21.40 & 22.54 & 24.11 & 24.30 
& 0.562 & 0.440 \\ 
  HDF3646+1237 & 12\ 36\ 46.96 & 62\ 12\  37.95 & 21.08 & 21.50 & 22.26 & 22.91 
& 0.321 & 0.175 \\ 
  HDF3641+1239 & 12\ 36\ 41.63 & 62\ 12\  39.82 & 25.48 & 25.51 & 25.53 & 28.13 
& 2.591 & 2.300 \\ 
  HDF3650+1240 & 12\ 36\ 50.12 & 62\ 12\  40.94 & 20.76 & 21.31 & 22.14 & 22.76 
& 0.474 & 0.550 \\ 
  HDF3644+1241 & 12\ 36\ 44.08 & 62\ 12\  41.66 & 23.16 & 23.91 & 24.16 & 24.21 
& 0.873 & 0.900 \\ 
  HDF3655+1246 & 12\ 36\ 55.49 & 62\ 12\  46.35 & 21.95 & 22.93 & 23.58 & 23.79 
& 0.790 & 0.750 \\ 
  HDF3650+1246 & 12\ 36\ 50.18 & 62\ 12\  46.92 & 21.45 & 22.89 & 24.58 & 26.60 
& 0.678 & 0.625 \\ 
  HDF3644+1248 & 12\ 36\ 44.11 & 62\ 12\  48.97 & 21.49 & 21.98 & 22.55 & 22.94 
& 0.558 & 0.625 \\ 
  HDF3649+1249$^8$ & 12\ 36\ 49.74 & 62\ 12\  49.94 & 25.17 & 25.44 & 26.62 & 28.01 
& 3.233 & 3.200 \\ 
  HDF3655+1250 & 12\ 36\ 55.57 & 62\ 12\  50.01 & 23.12 & 23.84 & 24.06 & 23.92 
& 0.950 & 0.925 \\ 
  HDF3658+1253 & 12\ 36\ 58.66 & 62\ 12\  53.26 & 21.50 & 21.95 & 22.76 & 23.17 
& 0.320 & 0.470 \\ 
  HDF3656+1253 & 12\ 36\ 56.66 & 62\ 12\  53.56 & 23.18 & 23.99 & 24.28 & 24.03 
& 1.231 & 0.900 \\ 
  HDF3647+1253 & 12\ 36\ 47.47 & 62\ 12\  53.68 & 23.74 & 24.36 & 25.12 & 25.03 
& 0.681 & 0.700 \\ 
  HDF3650+1257 & 12\ 36\ 50.73 & 62\ 12\  57.07 & 22.47 & 22.71 & 23.31 & 23.78 
& 0.319 & 0.240 \\ 
  HDF3647+1257$^{6,8}$ & 12\ 36\ 47.69 & 62\ 12\  57.35 & 24.87 & 25.26 & 26.24 & 28.01 
& 2.931 & 3.100 \\ 
  HDF3657+1301 & 12\ 36\ 57.99 & 62\ 13\   1.50 & 22.36 & 22.57 & 23.13 & 23.28 
& 0.319 & 0.220 \\ 
  HDF3648+1310 & 12\ 36\ 48.00 & 62\ 13\  10.07 & 20.54 & 21.58 & 23.28 & 25.22 
& 0.475 & 0.410 \\ 
  HDF3649+1312 & 12\ 36\ 49.29 & 62\ 13\  12.32 & 22.05 & 22.62 & 23.44 & 23.98 
& 0.478 & 0.575 \\ 
  HDF3649+1314 & 12\ 36\ 49.58 & 62\ 13\  14.31 & 21.58 & 22.39 & 23.59 & 24.36 
& 0.475 & 0.575 \\ 
  HDF3655+1315$^2$ & 12\ 36\ 54.92 & 62\ 13\  15.70 & 23.96 & 24.48 & 25.13 & 25.48 
& 0.511 & 0.650 \\ 
  HDF3654+1315 & 12\ 36\ 54.64 & 62\ 13\  15.76 & 23.95 & 24.09 & 24.33 & 24.56 
& 2.232 & 2.200 \\ 
  HDF3648+1319 & 12\ 36\ 48.75 & 62\ 13\  19.51 & 22.77 & 23.54 & 23.93 & 24.13 
& 0.749 & 0.800 \\ 
  HDF3651+1321 & 12\ 36\ 51.01 & 62\ 13\  21.31 & 20.13 & 20.46 & 21.05 & 22.08 
& 0.199 & 0.100 \\ 
  HDF3653+1322 & 12\ 36\ 53.17 & 62\ 13\  22.47 & 24.59 & 24.75 & 24.78 & 25.59 
& 2.489 & 2.300 \\ 
  HDF3648+1329 & 12\ 36\ 48.42 & 62\ 13\  29.55 & 23.05 & 24.03 & 24.33 & 25.25 
& 0.958 & 0.750 \\ 
  HDF3653+1330$^8$ & 12\ 36\ 53.34 & 62\ 13\  30.39 & 24.64 & 24.91 & 25.82 & 28.01 
& 2.991 & 3.000 \\ 
  HDF3652+1338$^8$ & 12\ 36\ 52.32 & 62\ 13\  38.73 & 24.97 & 25.33 & 26.55 & 28.01 
& 3.430 & 3.300 \\ 
  HDF3652+1340$^8$ & 12\ 36\ 52.66 & 62\ 13\  40.06 & 25.10 & 25.32 & 26.28 & 28.01 
& 3.368 & 3.100 \\ 
  HDF3654+1342$^8$   & 12\ 36\ 54.53 & 62\ 13\  42.29 & 25.32 & 25.31 & 25.46 & 27.93 
& 2.419 & 2.600 \\ 
  HDF3649+1347 & 12\ 36\ 49.35 & 62\ 13\  47.92 & 18.33 & 18.94 & 19.93 & 22.83 
& 0.089 & 0.075 \\ 
  HDF3545+1347$^7$   & 12\ 36\ 45.31 & 62\ 13\  47.92 & 24.91 & 24.99 & 25.49 & 26.73 
& 3.160 & 3.100 \\ 
  HDF3654+1348$^8$  & 12\ 36\ 54.97 & 62\ 13\  48.10 & 24.59 & 24.62 & 24.82 & 26.71 
& 2.233 & 2.275 \\ 
  HDF3655+1355 & 12\ 36\ 55.44 & 62\ 13\  54.58 & 22.63 & 23.26 & 23.48 & 23.66 
& 1.148 & 1.400 \\ 
  HDF3651+1354 & 12\ 36\ 51.69 & 62\ 13\  54.79 & 21.15 & 22.03 & 23.03 & 23.82 
& 0.557 & 0.650 \\ 
  HDF3652+1355 & 12\ 36\ 52.71 & 62\ 13\  55.52 & 22.03 & 22.33 & 22.44 & 22.92 
& 1.355 & 1.600 \\ 
  HDF3651+1401 & 12\ 36\ 51.94 & 62\ 14\   1.99 & 22.97 & 23.68 & 24.38 & 24.57 
& 0.557 & 0.700 \\ 
  HDF3650+1402 & 12\ 36\ 50.04 & 62\ 14\   2.10 & 24.26 & 24.63 & 24.59 & 25.50 
& 2.237 & 2.000 \\ 
  HDF3646+1405$^1$   & 12\ 36\ 46.26 & 62\ 14\   5.69 & 21.77 & 22.83 & 23.85 & 25.43 
& 0.960 & 0.625 \\ 
  HDF3652+1405 & 12\ 36\ 52.81 & 62\ 14\   5.77 & 23.16 & 24.01 & 24.73 & 24.54 
& 0.498 & 0.750 \\ 
  HDF3649+1407 & 12\ 36\ 49.43 & 62\ 14\   7.74 & 21.78 & 22.77 & 23.49 & 23.89 
& 0.752 & 0.750 \\ 
  HDF3646+1408$^3$  & 12\ 36\ 46.33 & 62\ 14\   8.67 & 23.47 & 23.92 & 24.30 & 24.85 
& 0.130 & 0.675 \\ 
  HDF3643+1410$^7$  & 12\ 36\ 43.98 & 62\ 14\  10.96 & 24.32 & 24.40 & 24.45 & 27.07 
& 2.267 & 2.400 \\ 
  HDF3653+1411$^{5,8} $  & 12\ 36\ 53.51 & 62\ 14\  11.12 & 24.63 & 24.97 & 26.29 & 28.01 
& 3.181 & 3.400 \\ 
  HDF3645+1413 & 12\ 36\ 45.80 & 62\ 14\  13.16 & 24.67 & 24.69 & 24.83 & 26.55 
& 2.427 & 2.100 \\ 
  HDF3649+1416 & 12\ 36\ 49.72 & 62\ 14\  16.25 & 23.29 & 23.76 & 23.90 & 24.61 
& 2.001 & 1.800 \\ 
  HDF3648+1417$^9$   & 12\ 36\ 48.25 & 62\ 14\  17.70 & 23.27 & 23.57 & 23.77 & 25.07 
& 2.008 & 2.008 \\ 
  HDF3650+1419 & 12\ 36\ 50.28 & 62\ 14\  19.72 & 23.11 & 23.87 & 24.23 & 24.15 
& 0.816 & 0.900 \\ 

\end{tabular}  
\begin{list}{}{}

\item[$^1$] This galaxy is close to HDF3646+1408 ($z_{spec}=0.13$). 
   An aperture of $1.2''$ diameter is used for this object.  
\item[$^2$] This galaxy is within $3''$ of another object (Lanzetta et al 1997). An 
   aperture of $1.2''$ is used for this galaxy. 
\item[$^3$] A spectroscopic redshift of 0.13 is found for this galaxy (Cowie 1997). 
   However, this object is close to another galaxy 
   (HDF3646+1405) with $z_{spec} =0.96$ 
   and lies within $ 1''$ of 
   a brighter source. It is likely that the spectroscopic redshift of this 
   galaxy is in error. Lanzetta et al (1997) estimate a photometric redshift 
   of 1.2 for this object compared to our value of 0.675 (when using $1.2'' $
   aperture diameter). 
\item[$^4$] This galaxy lies within $1''$ of a slightly brighter galaxy with no available
    redshift. Lanzetta et al (1997) estimate a photometric redshift of 0.76
    for this object compared to our estimate of 0.925. It is likely that the 
    spectroscopic redshift of this object is in error due to confusion with 
    an overlapping source.   
\item[$^5$] This galaxy is undetected in the U-band and shows a very red $B-I$ colour. 
    It is likely that intervening absorbers have affected its SED.   
\item[$^6$] This object shows a very complex morphology with a disk galaxy a few
     arcsec to its southwest (Lowenthal et al 1997). An aperture of $1.2''$
     diameter is used for this galaxy.   
\item[$^7$] An aperture of $1.2''$ diameter is used for this galaxy due to close
companions.   
\item[$^8$] These objects are identified as UV `drop-outs' in Lowenthal et al (1997). 
   We assume $U=28.01$ mag. 
   An effective aperture of $1.2''$ diameter is used.   
\item[$^9$] This galaxy had been given a wrong spectroscopic redshift of 2.845. Its
   spectroscopic redshift was subsequently measured to be 2.008
   (Lanzetta et al 1997; Dickinson 1998).  
\end{list}   
\end{table}

\begin{table}
\caption{Sensitivity of the {\it rms} scatter between the photometric
and spectroscopic redshifts on the EPS model parameters (i.e. template SEDs)}  
\begin{tabular}{lllclllll}

Model & IMF & $m_l$ & 
Number of & \multicolumn{4}{c}{$z_{form}$} & 
{\it rms} \\   
  &  & & Templates & 
E & Sp & Sb & Irr  &\\   
  &        &    &       &           &   &    &    &          \\ 
 1 & Salpeter  &      0.01 &      4  &  5&  5&    5 &   1 &     0.18 \\
 2 & "  &      0.01 &      3  &  5&  5&    5 &   -- &    0.26 \\ 
 3 & "  &      0.01 &      3  &  5&  2&    5 &   -- &    0.24 \\ 
 4 & "  &      0.01 &      4  &  5&  2&    5 &   1  &    0.11 \\ 
          &           &         &   &   &    &    &         \\ 
 5 & Salpeter  &      0.10 &      4  &  5&  5 &    5 &   1 &     0.27 \\ 
 6 & "       &      0.10 &      3  &  5&  5 &   5 &   -- &    0.38 \\ 
 7 & "       &      0.10 &      3  &  5&  2 &   5 &   -- &    0.37 \\ 
 8 & "       &      0.10 &      4  &  5&  2 &   5 &   1  &    0.22 \\ 
          &           &         &   &   &    &    &         \\  
 9 & Scalo     &      0.10 &      4  &  5&  5 &   5 &   1  &    0.28 \\ 
 10 & "     &      0.10 &      3  &  5&  5 &   5 &   -- &    0.52 \\ 
 11 & "     &      0.10 &      3  &  5&  2 &   5 &   -- &    0.45 \\ 
 12 & "     &      0.10 &      4  &  5&  2 &   5 &   1  &    0.25 \\ 
\end{tabular}   
\end{table}

\begin{table}
\caption{Changes in UV extinction ($A_U$) with redshift, corresponding to
 EPS models for Elliptical and Spiral galaxies.} 
\begin{tabular}{lllllll}  
    &   
\multicolumn{2}{c}{Salpeter IMF} &
\multicolumn{2}{c}{Salpeter IMF} & 
\multicolumn{2}{c}{Scalo IMF} \\
 & 
\multicolumn{2}{c}{$m_l =0.01$ M$_\odot$} & 
\multicolumn{2}{c}{$m_l =0.1$ M$_\odot$} & 
\multicolumn{2}{c}{$m_l =0.1$ M$_\odot$} \\
redshift &
{$A_U$(E)} & $A_U$(Sp) & $A_U$(E) & 
$A_U$(Sp) &
$A_U$(E) & $A_U$(Sp) \\  

        &         &          &         &        &        &       \\ 
   0    &     .02 &   0.78   &    .02  &  2.18  &  .02   & 2.45  \\
   0.5  &     .44 &   0.35   &    .34  &  1.22  &  .29   & 1.44  \\
   1    &    0.67 &   0.15   &    .56  &  0.57  &  .43   & 0.70  \\
   1.5  &    2.32 &   0.04   &    1.48 &  0.16  &  .83   & 0.20  \\
   2.   &    4.23 &     -    &    2.23 &     -  &  1.17  &    -  \\
   2.5  &    4.79 &     -    &    2.28 &     -  &  1.25  &    -  \\
\end{tabular}   
\end{table}

\vskip 3 truecm

\begin{table}   
\caption{    
Results from the simulation of spectral types.
Fraction of the galaxies of different types, correctly classified 
between the input and output simulated catalogues are listed.}
\begin{tabular}{lllllllll}  
     &
\multicolumn{4}{c}{Photometric Errors}  &
\multicolumn{4}{c}{Percentage$^1$}  \\    
     &
$\Delta(U)$    &  
$\Delta(B)$    &  
$\Delta(V)$    & 
$\Delta(I)$    & 
E    & 
Sp    & 
Sb    & Irr  \\  
              &     &     &     &     &       &      &     &     \\
simulation 1  & 0.15& 0.10& 0.05& 0.05& 100\% & 79\% & 71\%& 85\% \\
              &     &     &     &     &      &      &      &     \\
simulation 2  & 0.25& 0.15& 0.10& 0.10& 100\%& 66\% & 49\% & 62\% \\
              &     &     &     &     &      &      &      &     \\ 
\end{tabular} \vspace{-0.1in}   

\begin{list}{}{}
\item[$^1$] Ratio of the number of galaxies of a given spectral type in the
output catalogue which are correctly classified, to the total number of 
galaxies of the same spectral type in the input catalogue.    
 
\end{list}   
\end{table}  
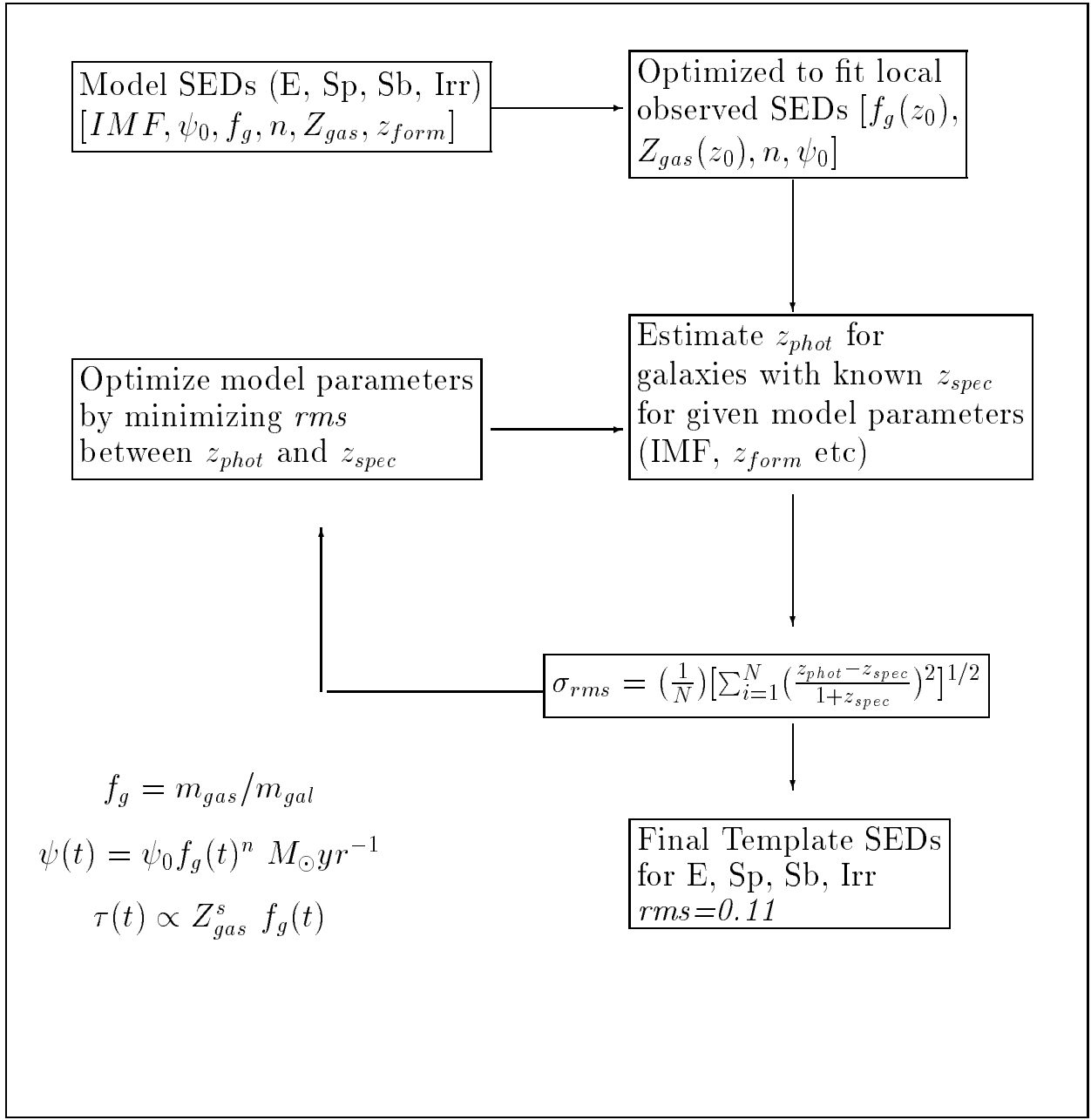
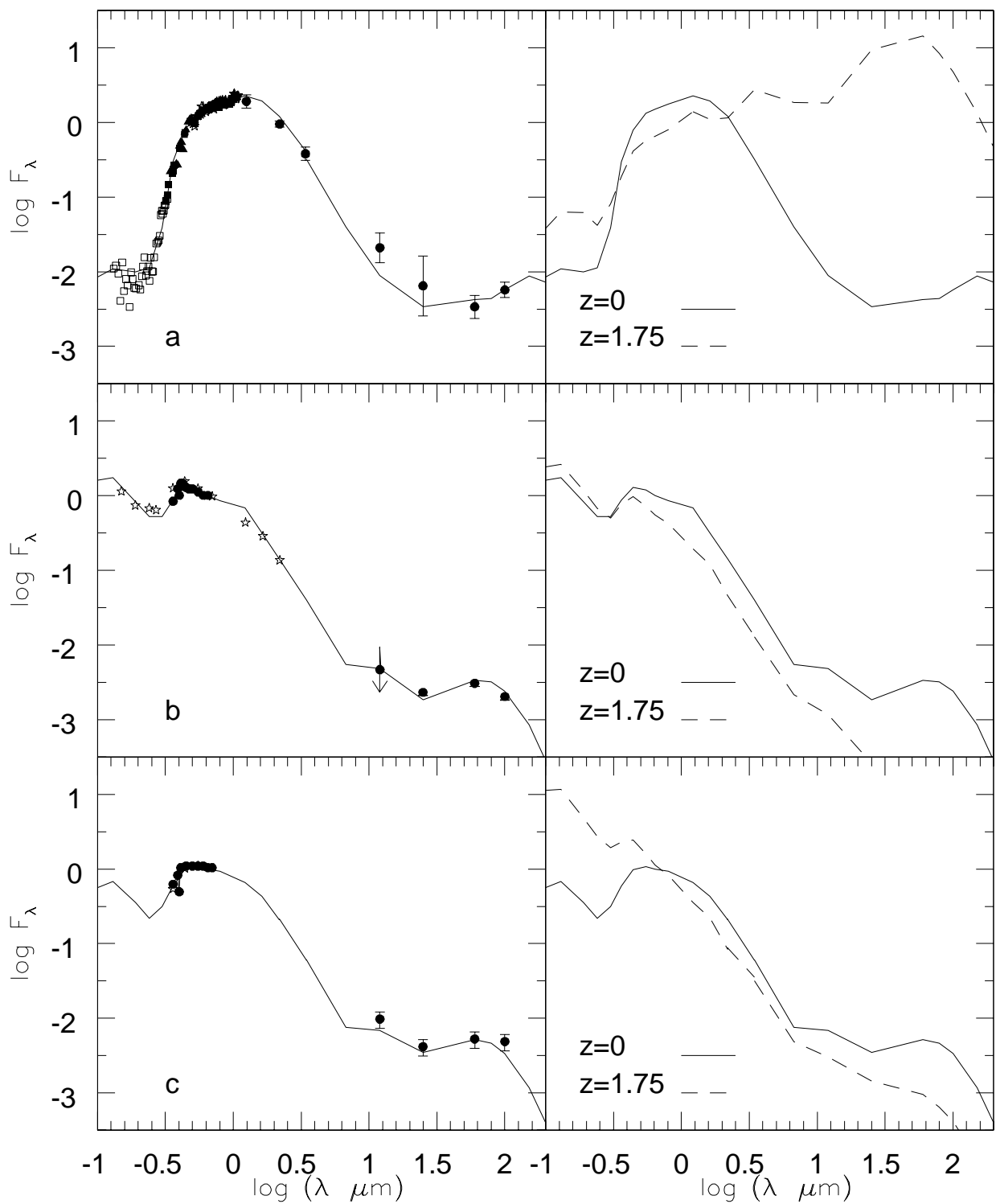
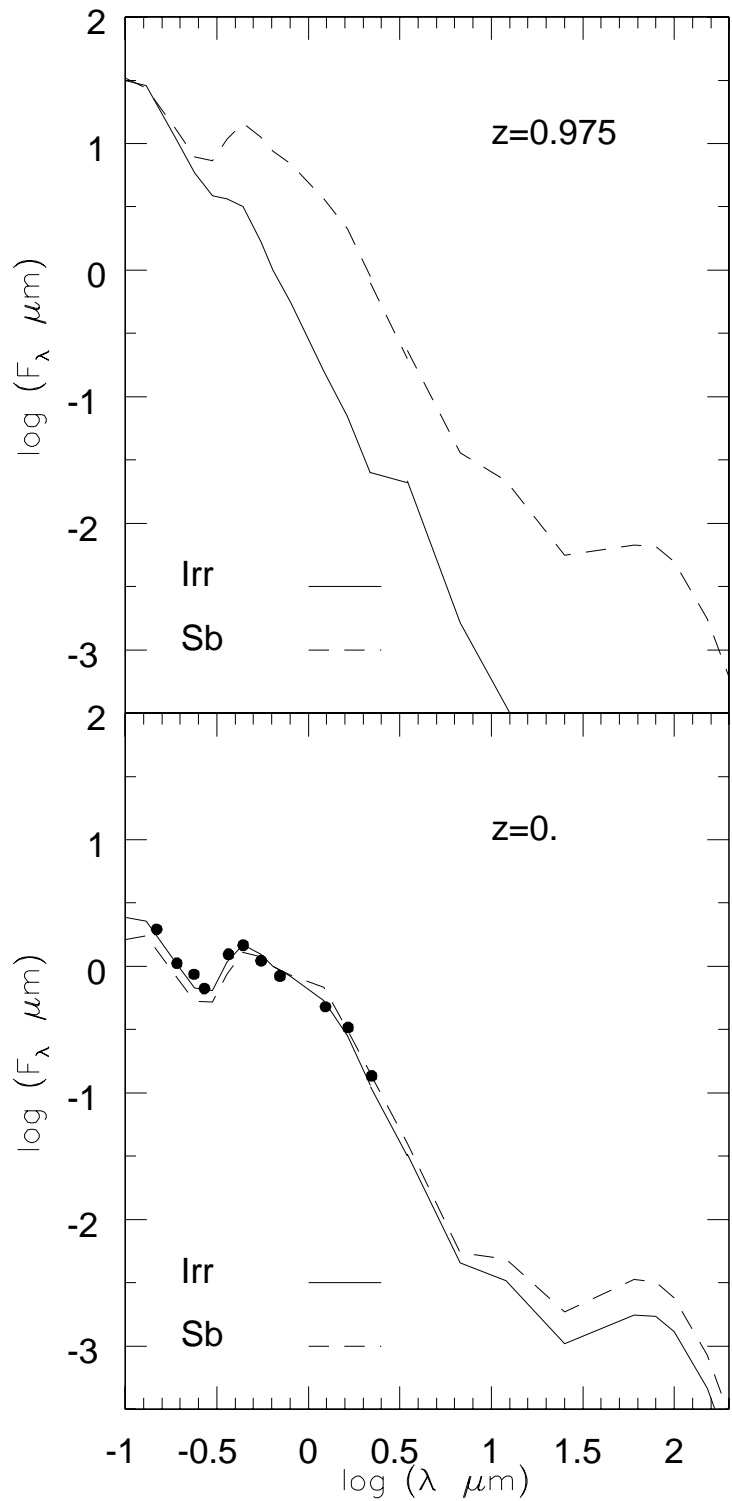
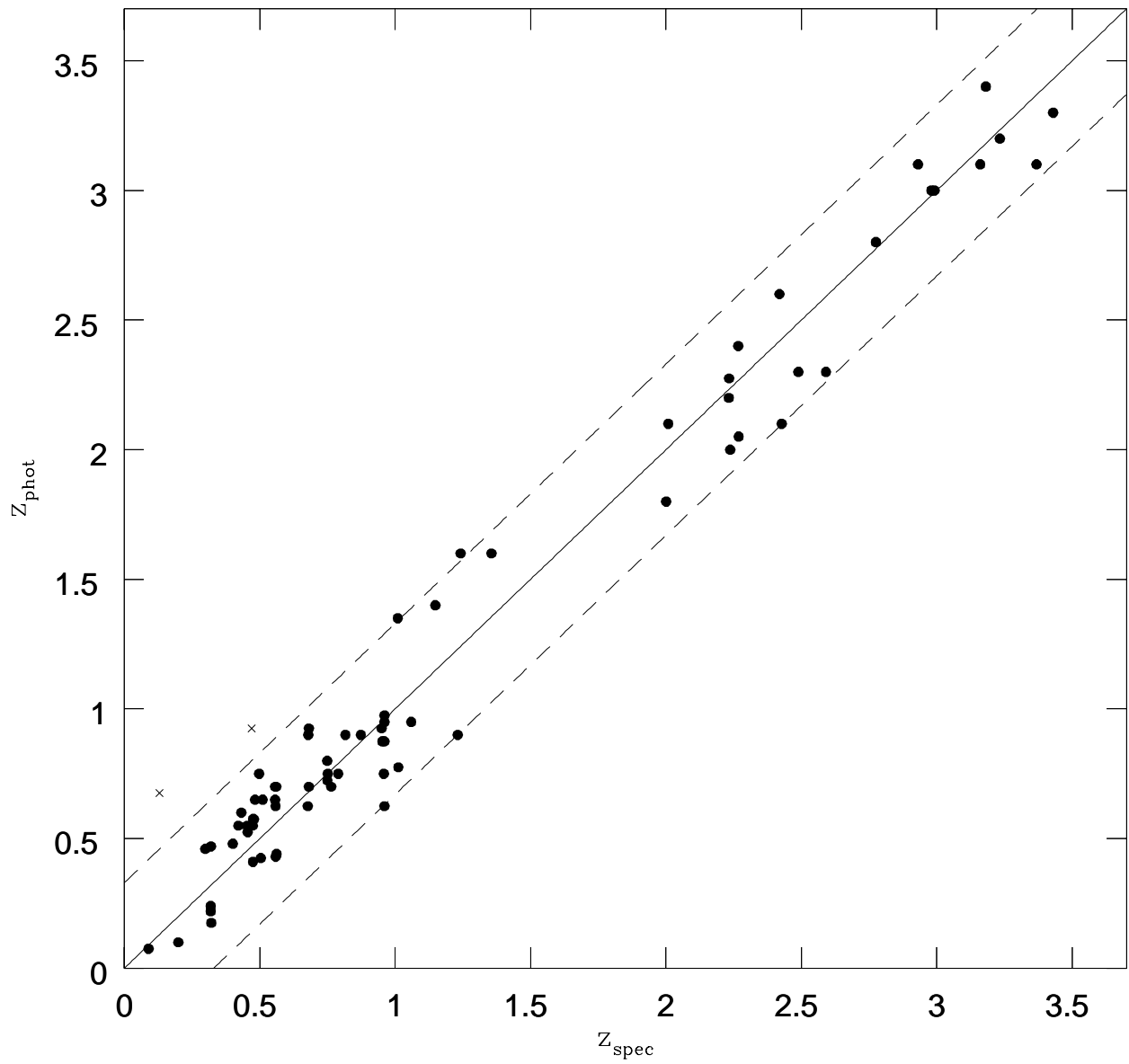
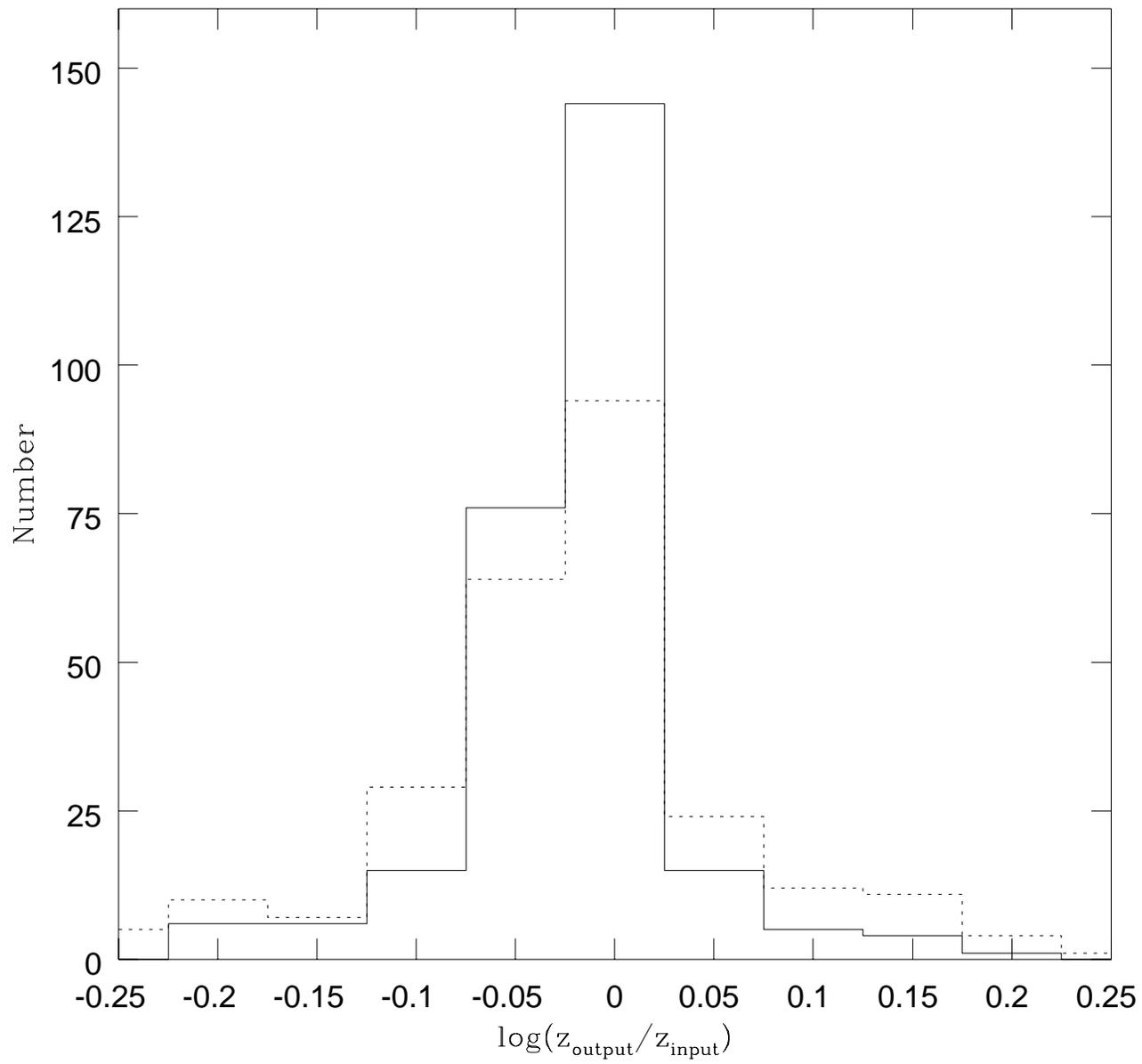
\end{document}